\documentclass[12pt,a4paper]{conference}

\usepackage{psfrag}
\usepackage{epsfig}
\usepackage{fancyhdr}
\usepackage{graphicx,amsmath,amssymb,cite}
\usepackage{multind}
\makeindex{author} \makeindex{subject}
\newcommand{\beq}{\begin{equation}}
\newcommand{\eeq}[1]{\label{#1}\end{equation}}
\newcommand{\eeqn}{\end{equation}}

\newcommand{\beqa}{\begin{eqnarray}}
\newcommand{\eeqa}[1]{\label{#1}\end{eqnarray}}
\newcommand{\eeqan}{\end{eqnarray}}

\let\bar=\overbar

\newcommand{\Dslash}{\not{\hbox{\kern-4pt $D$}}}
\newcommand{\dslash}{\not{\hbox{\kern-2pt $\del$}}}

\newcommand{\msb}{{\bar{\ssstyle M \kern -1pt S}}}

\begin{document}

\Chapter{Pion--deuteron scattering length in Chiral Perturbation  Theory up to order $\chi^{3/2}$}
{$\pi d$ scattering length in ChPT up to order $\chi^{3/2}$}{V. Baru \it{et al.}}
\vspace{-6 cm}\includegraphics[width=6 cm]{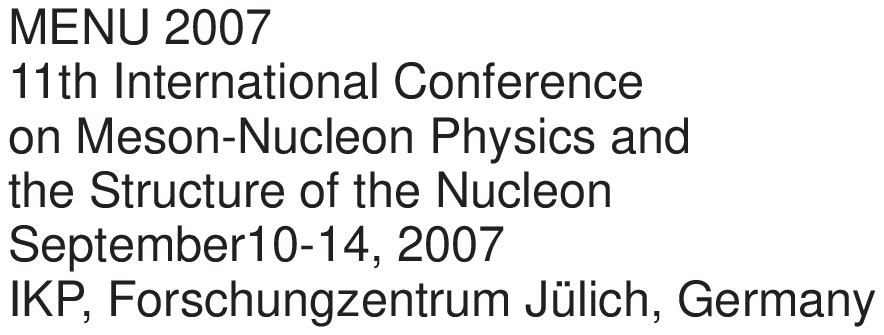}
\vspace{4 cm}

\addcontentsline{toc}{chapter}{{\it N. Author}} \label{authorStart}

\begin{raggedright}

\underline{V. Baru}$^{\star}$$^,$\footnote{E-mail address: baru@itep.ru}~, 
J. Haidenbauer$^{\%}$,  C.~Hanhart$^{\%}$,
A. Kudryavtsev$^{\star}$, V. Lensky$^{\star,\%}$ and U.-G. Mei\ss ner$^{\%,\#}$

\bigskip\bigskip


$^{\star}$Institute of Theoretical and Experimental Physics,
 117259, B.Cheremushkinskaya 25, Moscow, Russia \\
$^{\%}$Institut f\"{u}r Kernphysik, Forschungszentrum J\"{u}lich GmbH,
 D--52425 J\"{u}lich, Germany\\
$^{\#}$Helmholtz-Institut f\"{u}r Strahlen- und Kernphysik (Theorie), 
Universit\"at Bonn,  Nu{\ss}allee 14-16, D--53115 Bonn, Germany \\

\end{raggedright}


\begin{center}
\textbf{Abstract}
\end{center}

A complete calculation of the corrections to pion-deuteron scattering length up to order
$\chi^{3/2}$ with $\chi=m_{\pi}/M_{N}$ is performed. The calculation includes the dispersive 
contributions and corrections due to the explicit treatment of the
$\Delta$ resonance. s-wave pion-nucleon scattering parameters are extracted from
a combined analysis of modern experimental data.

\section{Introduction}

The pion-nucleon ($\pi N$) scattering lengths are fundamental
quantities of low--energy hadron physics since they test the QCD
symmetries and the pattern of chiral symmetry breaking. As stressed by
Weinberg long time ago, chiral symmetry suppresses the isoscalar $\pi
N$ scattering length $a^+$ substantially compared to its isovector
counterpart $a^- \, $. Thus, a precise determination of $a^+$ demands
in general high accuracy experiments.

Here
pion-deuteron ($\pi d$) scattering near threshold plays an exceptional role for
$\mbox{Re}(a_{\pi d}) = 2a^+ + (\mbox{few--body corrections}).$
 The first term $\sim a^+$
is simply generated from the impulse approximation (scattering off the proton
and off the neutron) and is independent of the deuteron structure.
Thus, if one is able to calculate the few--body corrections  in a controlled way, 
$\pi d$ scattering is a prime reaction to extract $a^+$
(most effectively in combination with pionic hydrogen measurements).

A method  how to calculate processes on few
nucleon systems with external probes was proposed by Weinberg in one
of his classical papers \cite{swein1}. As a first step the
perturbative transition operators need to be calculated using the
rules of ChPT. Then those transition operators must be convoluted with the appropriate $NN$ wave
functions. This scheme was already applied to a large number of
reactions like $\pi d\to \pi d$ \cite{beane}, $\gamma d\to \pi^0 d$
\cite{kbl,krebs}, $\pi {^3}$He$\to \pi {^3}$He \cite{baru}, $\pi^- d\to \gamma
nn$ \cite{garde}, and $\gamma d\to \pi^+ nn$ \cite{lensky}, where only the
most recent references are given.  The standard expansion parameter 
$\chi= m_{\pi}/M_N$,  where  $m_\pi\, (M_N)$ is the pion (nucleon) mass, was used in most 
of these references.

It was also Weinberg who calculated the leading order few body
corrections to the $\pi d$ system, the most important of which -- the
diagram when pion rescatters on two nucleons with the
Weinberg-Tomozawa (WT) vertices -- is almost as large as the 
experimental value for the $\pi d$ scattering length.  
The diagrams at leading order were calculated in the fixed center
kinematics, i.e. with  static nucleons. An accounting for the
nucleon recoils leads to potentially sizable corrections of order of
$\chi^{1/2}$ in the standard Weinberg counting. The  non-analyticity 
of this correction is related to the  few--body singularities that are
employed in some pion--few-nucleon diagrams as demonstrated in
Refs. \cite{recoils,lensky}. 
It was the main result of  Ref. \cite{recoils} that the
importance of the resulting effect of all recoil terms is directly connected to the Pauli principle for the nucleons in the intermediate state. In particular, if
the s-wave NN-state is not allowed by quantum numbers, which is fulfilled
in the $\pi d$ process, the net effect of the recoil correction
is to be small due to a cancellation of individually large terms. 
At next-to-leading (NLO) order there are basically the same diagrams as at LO 
but with subleading vertices. The calculation performed in Ref. \cite{beane} showed 
that the sum of diagrams contributing at NLO vanishes. 
Furthermore,  the solution  for $\{a^+,a^-\}$ was found in Ref. \cite{beane} from a common intersection of three bands corresponding to the shift and width of pionic hydrogen atom \cite{pipexp} and to the shift of pionic deuterium \cite{pidexp}. Due to a cancellation of terms at orders $\chi^{1/2}$ and $\chi$ the results for $\{a^+,a^-\}$ extracted in ChPT  \cite{beane} turned out to be quite similar to the phenomenological 
calculations \cite{bk,tle}.  
However, the recent measurement of the width of $\pi^- p$ atom with much better accuracy \cite{gotta} seriously changes the picture. The problem is that an intersection region of the three bands and thus a unique solution for $\{a^+,a^-\}$ does not exist anymore. This 
finding means that something important is missing in our understanding of the $\pi d$
system.  This could be isospin symmetry breaking (ISB) effects that were recently found in 
Ref.\cite{Akaki} to give a huge effect to the $\pi d$ scattering although with large 
uncertainty. Another possibility is that higher order effects to the transition operators could be important.  In this presentation we discuss both possibilities. The influence of ISB effects on the extraction of the s-wave $\pi N$ scattering lengths is considered in sec. 2.  In sec. 3 and 4 we discuss corrections to 
the $\pi d$ scattering length emerging at order N$^{3/2}$LO ($\chi^{3/2}$). At this order two new classes of diagrams start to contribute. One of them, the so-called dispersive correction due to the process $\pi d\to NN\to \pi d$ is the subject of sec. 3. In sec. 4 we discuss the effect of the $\Delta$ isobar as explicit degree of freedom. 
The main results are summarized in sec. 5. 

\vspace*{-0.3cm} 
\section{ISB effects and s-wave $\pi N$ scattering lengths}

Since the leading one-body contribution $(\sim a^+)$ to the $\pi d$
scattering length  is chirally suppressed the role
of ISB effects in this process becomes significant. For the  $\pi^- d$ system so far only leading ISB corrections were evaluated \cite{Akaki}.
They were found to give a very large effect of order of 40\% to the 
$\pi^- d$ scattering length.  In this section we would like to investigate the influence of this correction on the combined analysis 
of experimental data and thus on the s-wave $\pi N$ scattering lengths. 
To account for the ISB correction at leading order
we should replace $2a^+$ by $a_{\pi^-p}+a_{\pi^- n}$ in the expression for 
$a_{\pi d}$, which agrees to the former only, if isospin were
an exact symmetry. The expressions for the $\pi N$ amplitudes with inclusion of ISB effects were derived in Ref. \cite{Akaki}:  

\begin{eqnarray} \label{apiN} 
\nonumber
{{a}_{\pi^- p}}&=&{ a^+} + {a^-}+\frac{1}{4\pi (1+\chi)}
\left( \frac{4(m_\pi^2-m_{\pi^0}^2)}{F_\pi^2}\,{c_1}
-\frac{e^2}{2}\,(4{f_1}+f_2)\right), \\ 
{{a}_{\pi^- n}}&=&{ a^+} - {a^-}+\frac{1}{4\pi(1+\chi)}
\left( \frac{4(m_\pi^2-m_{\pi^0}^2)}{F_\pi^2}\,{c_1}
-\frac{e^2}{2}\,(4{f_1}-f_2)   \right).
\end{eqnarray}
Here $m_{\pi} (m_{\pi^0})$ is the charged (neutral) pion mass, $F_{\pi}=92.4$ MeV and 
$c_1$ and $f_1, f_2$ correspond to the strong and electromagnetic 
LECs. 
Whereas $f_2$ and $c_1$ are known more or less well ($f_2=-(0.97\pm
0.38)~{\rm GeV}^{-1}$ \cite{Gasser} and $c_1=-0.9^{+0.2}_{-0.5}~{\rm GeV}^{-1}$ \cite{ulfc1}) 
the value for  $f_1$ ($|f_1|\leq 1.4~{\rm GeV}^{-1}$) is very
uncertain -- a naive dimensional analysis was used in 
Ref.~\cite{Akaki} to fix the latter.  
At the same time it is $f_1$ and $c_1$ that give the largest
contribution to the ISB correction for $\pi d$ scattering thus introducing a large uncertainty in the extraction of  $\{a^+,a^-\}$ from the data. 
At this stage we would like to note that the parameters $a^+,  c_1$ and $f_1$  enter  the expressions for $a_{\pi^- p}$ and $a_{\pi^- n}$  in the same linear combination (see Eq.~(\ref{apiN})). Note that
the expression for the charge exchange amplitude 
$\pi^- p\to \pi^0 n$ does not depend on the LECs $c_1$ and $f_1$ at all. 
Therefore let us 
introduce  the quantity $\tilde a^+$ which is defined as
\begin{eqnarray}
\tilde a^+= a^+ + \frac{1}{4\pi(1+\chi)} \left(\frac{4(m_\pi^2-m_{\pi^0}^2)}{F_\pi^2}\,c_1 
- 2 e^2 f_1\right).
\end{eqnarray}
The leading  isospin breaking terms that are also the main source of the
uncertainty\footnote{The idea of using some linear combinations of observables to reduce the uncertainty was suggested in Refs~\cite{Akaki, AkakiTrento}. In particular the combination $2 a_{\pi^- p}-a_{\pi^- d}$ that depends solely on $a^-$ was considered in Ref.\cite{AkakiTrento}.}  are contained now in $\tilde a^+$. 
Using this the expressions for $a_{\pi^- p}$,  $a_{\pi^- n}$ and ${\rm Re}\,  a_{\pi d}$ take the form
\begin{eqnarray} \label{system} 
\nonumber
{{a}_{\pi^- p}}&=&{ \tilde a^+} + {a^-}-\frac{1}{4\pi (1+\chi)}
\frac{e^2}{2}\,f_2, \\ \nonumber
{{a}_{\pi^- n}}&=&{ \tilde a^+} - {a^-}+\frac{1}{4\pi(1+\chi)}
 \frac{e^2}{2}\,f_2   ,\\
{\rm Re}\, a_{\pi d}&=& 2 { \tilde a^+} + \left<\mbox{few--body corrections }(a^-)\right> \ ,
\end{eqnarray}
Thus, we get a system of three equations for $a_{\pi^- p}$, Re $a_{\pi d}$ and $a_{\pi^- p\to\pi^0 n}$
(the explicit expression for the latter is given in Ref.~\cite{Akaki}) to determine $\tilde a^+$ and 
$a^-$. Once this determination is performed and provided that new and less uncertain information about the
LECs $c_1$ and $f_1$ is available from elsewhere, one will be able to extract $a^+$ directly without doing the analysis of pionic data once again. 
Let us discuss Eq.~(\ref{system}) in more detail. Note that the equations for 
the hydrogen and deuterium shifts ($a_{\pi^- p}$ and ${\rm Re}\,  a_{\pi d}$ respectively) written in terms of $\tilde a^+$ and $a^-$ and including  ISB effects at leading order are very close to those obtained in the isospin symmetric case in  Ref. \cite{beane} in terms of 
$a^+$ and $a^-$. The difference basically consists in the term
proportional to $f_2$ for the pionic hydrogen shift that gives a relatively small effect. Thus the main modification due to 
the inclusion of ISB effects at leading order consists in the replacement of $a^+$ by $\tilde a^+$.
At the same time we know that  for the isospin symmetric case there is no unique solution for $\{a^+,a^-\}$ 
if the new data for the hydrogen width are utilized. Thus we conclude
that the system of equations for $\{\tilde a^+,a^-\}$ does not have a
unique solution either at least as long as the few body corrections of Ref.~\cite{beane} are 
used. In the following sections we present the recent progress in this sector.

\vspace*{-0.3cm}
\section{Dispersive corrections}

\begin{figure}[t!]
\begin{center}
\psfig{file=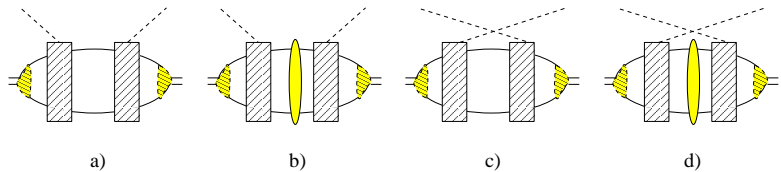,width=4.in}
\end{center}
\caption{Dispersive corrections to the $\pi d$ scattering length.}
\label{disp}
\end{figure}

Experimental  measurement of the $\pi d$ scattering length shows that its imaginary part is relatively large, about 1/4 of its real part \cite{pidexp}. 
The imaginary part can be expressed  in terms of the $\pi d$ total cross section through the
optical theorem. One gets
\begin{equation}
4\pi \mbox{Im}(a_{\pi d})=\lim_{q\to 0}q\left\{
\sigma(\pi d\to NN)+\sigma(\pi d\to \gamma NN)\right\} \ ,
\label{opttheo}
\end{equation}
where $q$ denotes the relative momentum of the initial $\pi d$ pair.
The ratio $R=\lim_{q\to 0}\left(\sigma(\pi d\to NN)/\sigma(\pi d\to
\gamma NN)\right)$ was measured to be $2.83\pm 0.04$
\cite{highland}. At low energies diagrams that lead to a sizable
imaginary part of some amplitude are expected to contribute also
significantly to its real part.
 Those contributions are called dispersive corrections.  As a
first estimate Br\"uckner speculated that the real and imaginary part
of these contributions should be of the same order of magnitude
\cite{brueck}. This expectation was confirmed within Faddeev
calculations in Refs. \cite{at}.  
Here we present the first consistent ChPT calculation of the 
dispersive corrections that emerge from the hadronic $\pi d\to NN\to \pi d$ 
and photonic $\pi d\to \gamma NN\to \pi d$ processes \cite{apiddisp}. 
We define dispersive corrections as contributions from diagrams 
with an intermediate state that contains only nucleons, photons and at most real pions.
Therefore, all the diagrams shown in Fig.~\ref{disp} are included in our work. All these diagrams contribute at order $\chi^{3/2}$ as compared to the leading double scattering diagram(see Ref.~\cite{apiddisp} for details).
The hatched blocks in the diagrams of Fig.~\ref{disp} refer
to the relevant transition operators for the reaction $NN\to NN\pi$ that consist 
of the direct and rescattering mechanisms. Note that the latter is to be calculated 
with the on--shell $\pi N\to \pi N$ vertices ($2m_\pi$) as was derived in Ref. \cite{NNpi}. Using the CCF potential~\cite{ccf}
for the $NN$ distortions we found for the dispersive
correction from the purely hadronic transition
\begin{equation}
\delta a_{\pi d}^{disp}=(-6.5+1.3+2.4-0.2)\times 10^{-3} \, m_\pi^{-1}
=-3.0\times 10^{-3} \, m_\pi^{-1} \ ,
\end{equation}
where the numbers in the first bracket are the individual results for the
diagrams shown in Fig.~\ref{disp}, in order.  Note that the diagrams with
intermediate $NN$ interactions and the crossed ones (diagram $(c)$ and $(d)$),
neither of them were included in most of the previous calculations, give
significant contributions. 
When repeating the calculation with the four different phenomenological 
$NN$ potentials CD Bonn \cite{cdbonn}, Paris \cite{paris}, AV18 \cite{argonne}  
we find
\begin{equation}
\delta a_{\pi d}^{\rm disp} = (-2.9\pm 1.4) \times
\,10^{-3} \ m_\pi^{-1} \ ,
\label{result_disp}
\end{equation}
where the first number is the mean value for the various potentials and the second number
reflects the theoretical uncertainty of this calculation estimated conservatively --- see Ref.~\cite{apiddisp} for details.  
Note that the same calculation
gave very nice agreement for the corresponding imaginary part~\cite{apiddisp}.

In Ref.~\cite{apiddisp} also the electromagnetic contribution to the
dispersive correction was calculated. It turned out that the contribution to
the real part was tiny --- $-0.1\times 10^{-3} \, m_\pi^{-1}$ --- while the
sizable experimental value for the imaginary part  was described well.

\vspace*{-0.3cm}
\section{Role of the Delta resonance}
From phenomenological studies it is well known that the delta isobar $\Delta(1232)$
plays a very special role in low energy nuclear
dynamics~\cite{ericsonweise} as a consequence of the
relatively large $\pi N\Delta$ coupling and the
 quite small
delta--nucleon mass difference $\Delta=M_\Delta-M_N\simeq 2m_\pi$,
where $M_\Delta$ denotes the mass of the delta.
In the present section we investigate the role of the $\Delta$ isobar in
the reaction $\pi d\to \pi d$ at threshold in EFT.  
In the delta-less theory the effect of the $\Delta$ resonance is hidden in the LEC $c_2$ which is the leading term in the so-called boost correction to the  $\pi d$ scattering length \cite{beane}. 
This correction is known to be quite sizable ($\sim$ 10-20\% of $a_{\pi d}$) although very model dependent. 
The pertinent one--body operator scales with the square of the nucleon momentum and therefore the corresponding expectation value is
proportional to the nucleon kinetic energy inside the deuteron ---
this quantity is strongly model-dependent~\cite{mitandreas}. 
However, the value of $c_2$ is reduced by
a large factor once the delta contribution is taken out \cite{BKM1,evgeninew} 
so that the residual boost correction becomes negligible \cite{delta}.

\begin{figure}[t!]
\begin{center}
\epsfig{file=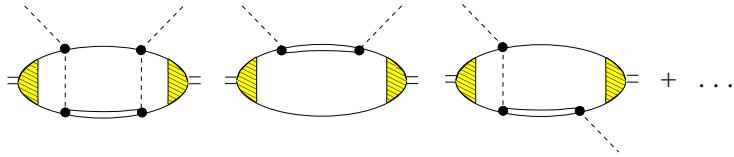, height=2.0cm,    angle=0} 
\caption{Diagrams with the $\Delta$ resonance at order $\chi^{3/2}$. Crossed terms are not shown explicitly but included in the calculation (see Ref.~\cite{delta}).
}
\label{diadelta}
\end{center}
\end{figure}

The reason why the explicit inclusion of the delta in pionic
reactions on the two--nucleon system is beneficial is that the dynamical treatment of the $\Delta$
allows to improve the convergence of the transition operators.
Let us, for example, focus on the one--body terms with the
delta 
(see, e.g., second diagram in Fig.~\ref{diadelta}). Then 
the corresponding $\pi N\to\pi N$ transition potential is proportional to $p^2/(m_\pi-\Delta-\vec
p\, ^2/M_N)$. 
  For static deltas, the nucleon-delta propagator reduces to
$1/(m_{\pi}-\Delta)$. 
Thus, in the latter case the
transition operator behaves like $\vec p\,^2$, whereas in the former
it approaches a constant for momenta larger than $|\vec p_\Delta \,
|\sim \sqrt{(\Delta-m_\pi)M_N}\sim 2.7 m_\pi$ with the effect that the
static amplitude is much more sensitive to the short range part of the
deuteron wave function and must be balanced by appropriate counter
terms.  The value of
$p_\Delta$ is numerically very close to $p_\mathrm{thr}=
\sqrt{M_Nm_\pi}\sim 2.6m_\pi$ --- the minimum initial momentum for the
reaction $NN\to NN\pi$ (for a recent review of this class of reactions
see Ref.~\cite{Hanrep}). 
This automatically puts the delta contributions in the same order as the dispersive
corrections 
\cite{delta}. 
In Fig. \ref{diadelta} we show 
diagrams with the $\Delta$ isobar that contribute at order
$\chi^{3/2}$. Diagrams with crossed external pions are not shown explicitly but are taken into account in the calculation. 
The resulting correction is \cite {delta}
\begin{equation}
\delta a_{\pi d}^\Delta = (2.38\pm 0.40) \times
\,10^{-3} \ m_\pi^{-1} \ ,
\label{result}
\end{equation}
where the central value is the arithmetic
average of the results for the seven different potentials
and the uncertainty reflects the variations in the
results. Note that we used phenomenological $NN$ models 
without \cite{paris,argonne,cdbonn} and with \cite{ccf} explicit delta degree of freedom, 
as well as three variants of $NN$ wave functions derived within
EFT \cite{evgeni}.
These numbers were obtained with the $\pi N\Delta$ coupling constant $h_A=2.77$.
In contrast to earlier treatments of the boost correction,
the results we found with the explicit treatment of the $\Delta$ depend only very weakly on the NN model used. 
In Ref.~\cite{delta} also a detailed comparison to previous
phenomenological works is given. 

\vspace*{-0.3cm}
\section{Results and Conclusions}
We performed a complete calculation of the isospin-conserving
corrections to the pion-deuteron scattering length up to order
$\chi^{3/2}$.  The calculation includes the dispersive 
contributions and corrections due to the dynamical treatment of the
$\Delta$ resonance. Although these corrections are quite significant
individually the net effect of the diagrams that contribute at order
$\chi^{3/2}$ is very small:
\begin{equation}
\delta a_{\pi d}^\Delta+\delta a_{\pi d}^{\rm disp}=(-0.6\pm 1.5)
\times \, 10^{-3} \ m_\pi^{-1}.
\end{equation}
However, an important consequence of our investigations is
that once the $\Delta$ is treated dynamically, as it is done here,
the so--called boost corrections contribute insignificantly 
to the $\pi d$ scattering length. 

\begin{figure}[t!]
\begin{center}
\epsfig{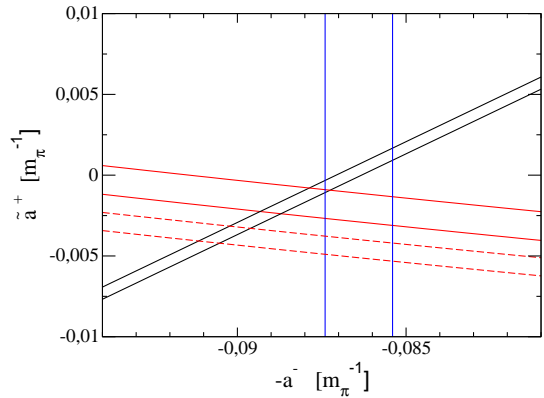} 
\caption{The solution for $\{\tilde a^+,a^-\}$ from the combined
analysis of experimental data. See text for details.
}
\label{b0b1}
\end{center}
\end{figure}

Also we analyzed the role of ISB effects at leading order in the
combined analysis of pionic data.  
It was observed that the LEC  $f_1$, that is known very poorly,
appears in the expressions for $a_{\pi^- p}$ and $a_{\pi^- n}$ in the
same linear combination with $a^+$ and the LEC $c_1$. We called it
$\tilde a^+$. Thus, the inclusion of ISB effects at leading order consists basically in the replacement of $a^+$ by $\tilde a^+$ in the combined
analysis of pionic data.   This 
drastically reduces the uncertainty of the analysis that
originates mainly from our ignorance regarding $f_1$.
The solution for the s-wave $\pi N$ parameters  $\{\tilde a^+,a^-\}$ 
is shown  in Fig.~\ref{b0b1}. The black
band stems from the analysis of the pionic hydrogen shift
\cite{gotta}. The blue vertical band corresponds to 
the new preliminary data for the
pionic hydrogen width \cite{gotta2}. The red solid and dashed  bands correspond
to the pion deuteron scattering length calculated with and without
corrections at order $\chi^{3/2}$. Also the boost correction was not
included in the full calculation corresponding to the solid red band
as a consequence of the explicit treatment of the $\Delta$ resonance. 
It is basically the latter effect that improves the situation resulting in some
intersection region for all three bands. 
However, it still remains to be seen if the corrections at NLO of
the isospin violation do not distort this picture. Corrections at this
order for the $\pi^- p$ system were evaluated in
Refs.~\cite{Gasser,nadia2} and turned out to be quite sizable,
especially those that come from the pion mass difference.  In order to
push also the calculation for the $\pi d$ system to a similar level of
accuracy in isospin violation, the $\pi^- n$ scattering amplitude as
well as some virtual photon exchanges in the $\pi^- d$ system are
still to be calculated. 
\vspace*{-0.5cm}
\section*{Acknowledgments}
This research is part of the EU Integrated Infrastructure Initiative
Hadron Physics Project under contract number RII3-CT-2004-506078, and
was supported also by the DFG-RFBR grant no. 05-02-04012 (436 RUS
113/820/0-1(R)) and the DFG SFB/TR 16 "Subnuclear Structure of Matter".  V.~B and
A.~K. acknowledge the support of the Federal Agency of Atomic Research
of the Russian Federation.

\vspace*{-0.5cm}


\begin{thebibliography}{0}    




\bibitem{swein1}
S.~Weinberg, Phys.  Lett.  B {\bf 295}, 114 (1992).

\bibitem{beane}
S.~R.~Beane, V.~Bernard, E.~Epelbaum, U.-G.~Mei{\ss}ner and D.~R.~Phillips,
Nucl.\ Phys.\ A {\bf 720}, 399 (2003) 
[arXiv:hep-ph/0206219].


\bibitem{kbl}S.~R.~Beane, V.~Bernard, T.~S.~H.~Lee, U.-G.~Mei{\ss}ner and U.~van Kolck,
  Nucl.\ Phys.\ A {\bf 618}, 381 (1997)
  [arXiv:hep-ph/9702226].

\bibitem{krebs} H.~Krebs, V.~Bernard and U.-G.~Mei\ss ner, 
Eur.\ Phys.\ J.\ A {\bf 22}, 503  (2004)  [arXiv:nucl-th/0405006].


\bibitem{baru}
  V.~Baru, J.~Haidenbauer, C.~Hanhart and J.~A.~Niskanen,
  Eur.\ Phys.\ J.\ A {\bf 16}, 437 (2003)
  [arXiv:nucl-th/0207040].



\bibitem{garde}
  A.~G{\aa}rdestig and D.~R.~Phillips,
  Phys.\ Rev.\ C {\bf 73}  014002 (2006);
 Phys.\ Rev.\ Lett.\  {\bf 96}, 232301 (2006),  [arXiv:nucl-th/0603045].


\bibitem{lensky}
  V.~Lensky, V.~Baru, J.~Haidenbauer, C.~Hanhart, A.~E.~Kudryavtsev and U.-G.~Mei\ss ner,
  Eur.\ Phys.\ J.\ A {\bf 26}, 107 (2005)
  [arXiv:nucl-th/0505039].


\bibitem{recoils}
  V.~Baru, C.~Hanhart, A.~E.~Kudryavtsev and U.-G.~Mei\ss ner,
  Phys.\ Lett.\ B {\bf 589}, 118 (2004)
  [arXiv:nucl-th/0402027].


\bibitem{pipexp}
H.~C.~Schr\" oder {\it et al.},
Eur.\ Phys.\ J.\  C {\bf    21}, 473 (2001).

\bibitem{pidexp}
 P.~Hauser {\it et al.},
  Phys.\ Rev.\  C {\bf 58}, 1869 (1998).


\bibitem{bk} V.~Baru, A.~Kudryavtsev, Phys. Atom. Nucl., {\bf 60},
  1476 (1997).

\bibitem{tle}
T.~ E.~ O.~ Ericson, B.~ Loiseau, A.~ W.~ Thomas, Phys. Rev. C {\bf 66}, 014005 (2002) [arXiv:hep-ph/0009312].


\bibitem{gotta}
  D.~Gotta,  
  Int.\ J.\ Mod.\ Phys.\  A {\bf 20}, 349 (2005).

\bibitem{Akaki}
  U.-G.~Mei\ss ner, U.~Raha and A.~Rusetsky,
  Phys.\ Lett.\  B {\bf 639}, 478 (2006)
  [arXiv:nucl-th/0512035].



\bibitem{Gasser}
  J.~Gasser, M.~A.~Ivanov, E.~Lipartia, M.~Mojzis and A.~Rusetsky,
  Eur.\ Phys.\ J.\  C {\bf 26}, 13 (2002)
  [arXiv:hep-ph/0206068].


\bibitem{ulfc1}
  U.-G.~Mei\ss ner,
  PoS {\bf LAT2005}, 009 (2006)
  [arXiv:hep-lat/0509029].

\bibitem{AkakiTrento}

 A.~Rusetsky, Talk given at the International Workshop on Exotic
 Hadronic Atoms, Deeply Bound Kaonic Nuclear States and Antihydrogen,
 Trento, Italy, 19-24 Jun 2006;  [arXiv: hep-ph/0610201]


\bibitem{highland}
V.~C.~Highland et al., Nucl.\ Phys.\ A {\bf 365}, 333 (1981).


\bibitem{brueck}
K.~Br\"uckner, Phys. Rev. {\bf 98}, 769 (1955).

\bibitem{at}
I.R.~Afnan and A.W.~Thomas, Phys. Rev. C{\bf 10}, 109 (1974);
 D.S.~Koltun and T.~Mizutani, Ann. Phys. (N.Y.) {\bf 109}, 1  (1978).

\bibitem{apiddisp}  V.~Lensky, V.~Baru, J.~Haidenbauer, C.~Hanhart, A.~E.~Kudryavtsev and U.-G.~Mei\ss ner,
  Phys.\ Lett.\  B {\bf 648}, 46 (2007)
  [arXiv:nucl-th/0608042].

\bibitem{NNpi}
  V.~Lensky, V.~Baru, J.~Haidenbauer, C.~Hanhart, A.~E.~Kudryavtsev and U.-G.~Mei\ss ner,
  Eur.\ Phys.\ J.\ A {\bf 27}, 37 (2006)
  [arXiv:nucl-th/0511054].

\bibitem{ccf}  J. Haidenbauer, K. Holinde, M.B.Johnson,
  Phys. Rev. C {\bf 48}, 2190 (1993).

\bibitem{cdbonn}
  R.~Machleidt, 
  Phys.\ Rev.\ C {\bf 63}, 024001 (2001)
  [arXiv:nucl-th/0006014].

\bibitem{paris}   M. Lacombe et al., Phys. Rev. C {\bf 21}, 861 (1980).

\bibitem{argonne}    R.~B.~Wiringa, V.~G.~J.~Stoks and R.~Schiavilla,
  Phys.\ Rev.\  C {\bf 51}, 38 (1995) 
  [arXiv:nucl-th/9408016].   

\bibitem{ericsonweise}
T. Ericson und W. Weise, {\em Pions and Nuclei} (Clarendon Press, Oxford,
  1988).

\bibitem{mitandreas}
  A.~Nogga and C.~Hanhart,
  Phys.\ Lett.\ B {\bf 634}, 210 (2006)
  [arXiv:nucl-th/0511011].

\bibitem{BKM1}
 V.~Bernard, N.~Kaiser and U.-G.~Mei\ss ner,
  Nucl.\ Phys.\  A {\bf 615}, 483 (1997) 
  [arXiv:hep-ph/9611253].

\bibitem{evgeninew}
  H.~Krebs, E.~Epelbaum and U.-G.~Mei\ss ner,
  Eur. \ Phys. \ J. A {\bf 32}, 127 (2007)
  [arXiv:nucl-th/0703087].

\bibitem{delta}
 V.~Baru, J.~Haidenbauer, C.~Hanhart, A.~Kudryavtsev, V.~Lensky and U.-G.~Mei\ss ner,
  arXiv:0706.4023 [nucl-th], Phys. Lett. B, in print.

\bibitem{Hanrep}
C.~Hanhart, {\it Phys. Rep.} \textbf{397}, 155 (2004) [arXiv:hep-ph/0311341].

\bibitem{evgeni}  E.~Epelbaum, W.~Gl\"ockle and U.-G.~Mei\ss ner,
  Nucl.~Phys.~A {\bf 747}, 362  (2005) [arXiv:nucl-th/0405048].

\bibitem{gotta2}   T.Strauch, 
Talk given at 
MENU 2007, Julich, Germany, 10-14 Sep 2007.


\bibitem{nadia2}
  N.~Fettes and U.-G.~Mei\ss ner,
  Nucl.\ Phys.\  A {\bf 693}, 693 (2001)
  [arXiv:hep-ph/0101030].





\end{thebibliography}
\end{document}